\documentclass[apj]{emulateapj}
\usepackage{comment}
\usepackage{color}
\usepackage[bookmarks,bookmarksnumbered,colorlinks=true, citecolor=blue, linkcolor=black]{hyperref}

\def \BE{\begin{equation}}
\def \EE{\end{equation}}	
\def \BC{\begin{center}}
\def \EC{\end{center}}
\def \BEA{\begin{eqnarray}}
\def \EEA{\end{eqnarray}}

\def \MSUN{\rm M_{\odot}}

\def \RVIR{R_{\rm vir}}

\def\KMS{\,{\rm km\,s^{-1}}}
\def\spose#1{\hbox to 0pt{#1\hss}}
\def\lta{\mathrel{\spose{\lower 3pt\hbox{$\mathchar"218$}}
     \raise 2.0pt\hbox{$\mathchar"13C$}}}
\def\gta{\mathrel{\spose{\lower 3pt\hbox{$\mathchar"218$}}
     \raise 2.0pt\hbox{$\mathchar"13E$}}}

\shorttitle{In-Situ vs. Ex-Situ Star Formation}
\shortauthors{Pillepich, Madau \& Majer}

\begin{document}

\title{Building Late-Type Spiral Galaxies by In-Situ and Ex-Situ Star Formation}

\author{Annalisa Pillepich$^{1,2}$, Piero~Madau$^{2}$, \& Lucio Mayer$^{3}$}

\affil{$^1$Harvard--Smithsonian Center for Astrophysics, 60 Garden Street, Cambridge, MA 02138, USA\\
$^2$Department of Astronomy \& Astrophysics, University of California Santa Cruz, 1156 High St., Santa Cruz, CA 95064, USA \\
$^3$Center for Theoretical Astrophysics and Cosmology, Institute for Computational Science, University of Zurich, Winterthurerstrasse 190, CH-9057 Zurich, Switzerland}


\begin{abstract}
We analyze the formation and evolution of the stellar components in ``Eris", a 120 pc-resolution cosmological hydrodynamic simulation of a late-type spiral galaxy.
The simulation includes the effects of a uniform UV background, a delayed-radiative-cooling scheme for supernova feedback, and a star formation recipe 
based on a high gas density threshold. It allows a detailed study of the relative contributions of ``in-situ" (within the main host) and ``ex-situ" (within 
satellite galaxies) star formation to each major Galactic component in a close Milky Way analog.
We investigate these two star-formation channels as a function of galactocentric distance, along different 
lines of sight above and along the disk plane, and as a function of cosmic time.  We find that: 1) approximately 70 percent of today's stars formed in-situ; 
2) more than two thirds of the ex-situ stars formed within satellites {\it after} infall; 3) the majority of ex-situ stars are found today in the disk and in the bulge; 
4) the stellar halo is dominated by ex-situ stars, whereas in-situ stars dominate the mass profile at distances $\lta 5$ kpc from the center at high latitudes; and 
5) approximately 25\% of the inner, $r\lta 20$ kpc, halo is composed of in-situ stars that have been displaced from their original birth sites during Eris' early assembly history.
\end{abstract}

\section{Introduction} \label{SEC:INTRO}

Knowledge of the spatial distribution, kinematics, and chemical composition of stellar populations in the Milky Way galaxy
provides invaluable information on the nature of its formation and evolution. In the standard $\Lambda$CDM scenario, 
hierarchical clustering leads to complex galaxy assembly histories, and the paths leading to the build up of each 
major stellar component may be different. 
The thin disk naturally forms from the dissipative infall of gas and angular momentum conservation. But old stars in the thick disk
may be brought in by satellites \citep{Abadi:2003} or result from the dynamical heating of the thin stellar disk by minor mergers 
\citep[e.g.,][]{Kazantzidis:2008}. The bulge of the Milky Way has signatures of both ``classical" and ``pseudo-bulges". Classical 
bulges are thought to form early, before the present disks were actually assembled, 
and rapidly, perhaps following a major merger with a satellite \citep{Hernquist:1991}. They may have composite stellar populations made of the
stars acquired from the infalling satellites and the newly created stars in the starburst triggered by the mergers \citep[e.g.,][]{Zavala:2012}. 
Pseudo-bulges, on the other hand, may grow slowly through internal secular processes that heat the inner disk \citep{Kormendy:2004,Athanassoula:2005,Debattista:2006} 
or form quickly at high redshift via a combination of non-axisymmetric disk instabilities and tidal interactions or mergers \citep{Guedes:2013}. Finally, the Galactic stellar 
halo may form from the disruption of accreted, lower-mass satellites \citep[e.g.,][]{Searle:1978,Bullock:2005,Abadi:2006,Bell:2008, Pillepich:2014b}. 
The same accretion and merger events are thought to be responsible for the displacement of stars formed within the bulge and the thin disk and found at tens of kpc distance 
from the Galactic center: these are the so called disk-heated stars, kicked-out disk stars, or ``in-situ halo stars'', the existence of which has been inferred from observations 
\citep[][]{Carollo:2010, Nissen:2010, Dorman:2014} and predicted by simulations \citep[e.g][]{Zolotov:2009, Purcell:2010}.

One basic structural distinction to be made is therefore between stars formed ``in situ", i.e., from gas condensing within the innermost 
regions of the main host, and those formed ``ex-situ", i.e., in some progenitor satellite dwarf of the parent galaxy. 
This grouping may be key, e.g., to understand stellar metallicity gradients, abundance ratios, kinematics, ages, degree of substructure, 
and the importance of dissipative processes in the assembly of each major Galactic component.
Recent theoretical studies of the relative contributions of these different star-formation channels have come mostly in two flavors: either via a combination of N-body simulations 
with semi-analytic models and/or stellar tagging techniques \citep[particle tagging for ex-situ stars only; ][]{Cooper:2010, Cooper:2013}; or via gas dynamics simulations of individual 
highly-resolved galaxies \citep{Abadi:2006, Zolotov:2009, Tissera:2012, Tissera:2013, Tissera:2014, Marinacci:2014} and of large samples of galaxies \citep{Oser:2010,Font:2011,McCarthy:2012, 
Lackner:2012}. Some of these works have prevalently focused on the galactic stellar halos. Nevertheless, a comprehensive and coherent picture is still elusive and a series of issues are 
still pressing: What is the relative importance of accreted, in-situ and satellite stars as a function of distance from the galactic center? How does this balance depend on halo mass? What 
physical mechanisms are responsible for the displacement of stars within and between its different morphological components? How many satellites contributed to the accreted stellar halo mass, when 
were they accreted, what were their properties before infall? Are there differences between the stars belonging to surviving self-bound satellites and the debris stars of disrupted accretion events? 
Are the lowest metallicity stars accreted or formed in-situ? Where do we expect to find the oldest stars, in globular clusters, satellite dwarf spheroidals, the halo, or the bulge? Are the 
most metal-poor stars also the oldest? And how do these predictions depend on the specific merger and star formation histories of individual galaxies?  

In this paper, we analyze ``Eris", one of the highest resolution, smoothed particles hydrodynamics (SPH) simulations ever run of a Milky Way-like galaxy formed from cosmological initial 
conditions \citep{Guedes:2011}. We investigate the contribution of in-situ and ex-situ stars, first simply as a function of distance and then by distinguishing among the different 
morphological components. We follow the assembly history of the main host and the two star-formation channels, and keep track of different sub-classes of ex-situ stars to properly 
describe the complexity of the phenomena under study. Our operational distinction between in-situ and ex-situ stars is meant to properly reflect the chemical and physical properties of 
the interstellar medium (ISM) out of which stars formed. The high mass, spatial, and time resolution of Eris are unprecedented in this type of analyses. Eris is the first simulated galaxy to satisfy a series of observational constraints for the Milky Way in terms of mass budget in the various components, structural properties, and scaling relations between mass and luminosity.
In particular, the combination of resolution and subgrid prescriptions allows to resolve the giant cloud complexes where star formation 
actually occurs in the ISM, to prevent artificial fragmentation, and to avoid the spurious formation of stars in small gas overdensities at large distances from the natural, dense sites 
of star-formation activity. All this makes Eris an excellent and reliable tool to understand the interplay between in-situ and ex-situ star formation, and a plausible 
high-resolution benchmark to aid the comparison across theoretical results and the interpretation of Milky Way's observations.

The plan of the paper is as follows. We introduce the specifics of the simulation, the adopted techniques, and the definitions of in-situ and ex-situ stars in Section 
\ref{SEC_METHODS}. In Section \ref{SEC_ASSEMBLY} we outline the assembly history of our simulated Milky Way analog. Our main results on the interplay between in-situ 
and ex-situ stars across the main host are given in Section \ref{SEC_ACROSSMW}. The spatial, kinematics, and chemical properties of in-situ and ex-situ stars 
are quantified and discussed in Section \ref{SEC_PROPERTIES}. We briefly discuss our findings in Section \ref{SEC_DISC} and summarize in Section \ref{SEC_FINAL}.

\begin{figure*}
\begin{center}
\includegraphics[width=17cm]{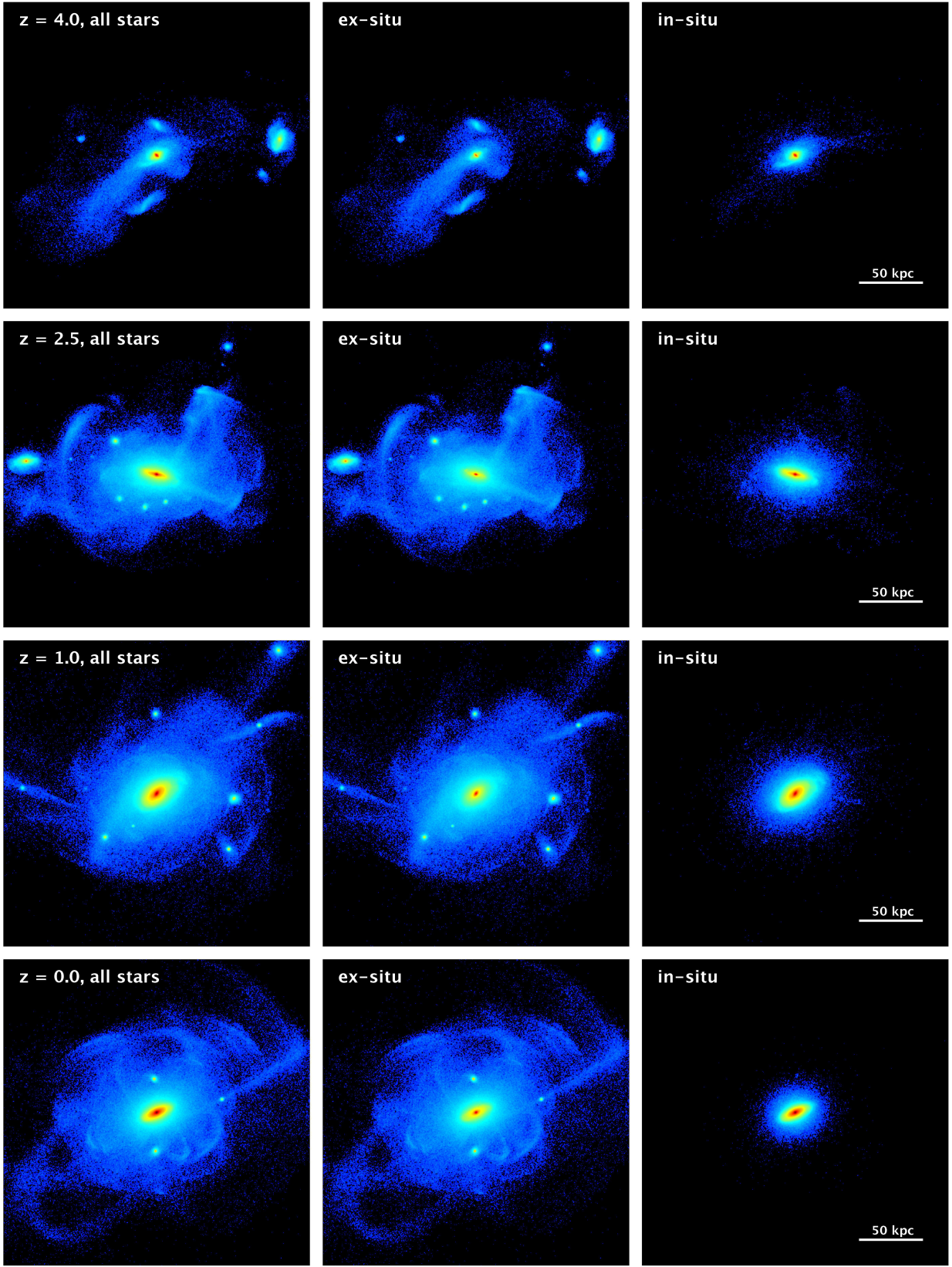}
\caption{Random projection Eris's stellar density field in a (240 comoving kpc)$^3$ box, from $z \sim 4$ (top) to today (bottom). The middle and right columns show only 
stars formed ex-situ and in-situ, respectively. Surface mass densities range from 100 to $10^8 \MSUN$kpc$^{-2}$. 
}
\label{FIG_MWDENSITY_EVOLUTION} 
\end{center}
\end{figure*}

\section{Methods and Definitions} \label{SEC_METHODS}

\subsection{Simulation}

Eris \citep{Guedes:2011} is a cosmological zoom-in simulation of a $M_{\rm vir}=8\times 10^{11}\MSUN$ galaxy halo followed from $z=90$ to the present epoch using the parallel 
TreeSPH code \textsc{Gasoline} \citep{Wadsley:2004a}. It includes Compton cooling, atomic cooling, and metallicity-dependent radiative cooling at low temperatures. 
A uniform UV background modifies the ionization and excitation state of the gas and is implemented using a modified version of the \citet{Haardt:1996} spectrum. 
The target halo was selected to have a quiet recent merger history, i.e., to have had no major mergers (mass ratio $\ge$ 1:10) after $z = 3$.
The Mpc-size high resolution region was resampled with 13 million dark matter
particles and an equal number of gas particles, for a mass resolution of $m_{\rm DM}=9.8\times 10^4\,\MSUN$ and $m_{\rm SPH}=2\times 10^4\,\MSUN$. The gravitational softening
length was fixed to 124  physical pc for all particle species from $z=9$ to the present time, and evolved as $1/(1+z)$ at earlier times. Star particles form in cold gas that
reaches a density threshold of $n_{\rm SF}=5$ atoms cm$^{-3}$, and are created stochastically with an initial mass $m_*=6\times 10^3\,\MSUN$ distributed following a \citet{Kroupa:1993}
initial mass function. Supernova (SN) explosions deposit an energy of $8\times 10^{50}\,$ergs and metals into a ``blastwave radius", and the heated gas has its radiative cooling delayed
following \citet{Stinson:2006}. The use of a high threshold for star formation has the effect of increasing the efficiency of SN feedback through the injection of energy in
localized high-density regions.

\subsection{Halo Finder and Stellar Tracking}

We have identified the main halo and its subhalos using the spherical overdensity Amiga Halo Finder \citep[{\tt AHF},][]{Gill:2004a,Knollmann&Knebe:2009}. 
The {\tt AHF} uses all particles (dark matter, gas, and stars) to find the centers of halos and subhalos as peaks in the particle density distribution. Spherical symmetry is then 
assumed to define the boundary of an isolated halo, $\RVIR$, as the point where the mean average density drops below the overdensity $\Delta_{\rm vir}$ (equal at $z=0$ to 
364 times the cosmological mean background density). A procedure to remove gravitationally unbound particles is recursively applied based on energy criteria. 
Effectively, the virial radius of {\tt AHF} subhalos is defined as the distance of the most distant bound particle to the center of the subhalo. Here, the center of a (sub)halo is chosen 
as the center of mass of all its bound particles. 

We have run the {\tt AHF} on 400 snapshots between $z=90$ and $z=0$ spaced at time intervals of $\sim$ 30 Myr. This allows us to follow reliably the merger histories of individual subhalos 
as well as the timelines of stars in the simulation. A catalog of all the luminous satellites ever fallen into the main host was compiled that is complete since redshift 10 and for 
stellar masses larger than $M_*=5\times 10^{4}\MSUN$ at accretion. Halo infall occurs when the center of a future satellite crosses for the first time the virial radius of the main host. 
At every snapshot, we flag every star of the simulation with the ID of the {\it smallest bound halo} which it is bound to, according to the {\tt AHF} catalog at that snapshot. 
For each star, we record its {\it formation time} (when the star particle forms out of a gas particle), {\it accretion or infall time} (when its host subhalo enters the Milky Way's virial 
radius), and {\it stripping time} (when the star is stripped from its host after infall). Formation times, easily convertible in stellar ages, are sampled 
more frequently than snapshot outputs, and have an uncertainty of half-a-million years. Accretion and stripping times depend on the frequency of snapshot outputs, 
with the stripping time of a star being by construction always larger or equal to its accretion time. The orbital properties of the subhalo host at accretion 
and the energy of its stars determine the difference between accretion and stripping times, which can be as large as a few Gyrs. Stars that form after infall in an orbiting 
satellite are assigned an accretion time equal to their formation time.

\subsection{In-Situ vs. Ex-Situ}
\label{SEC_DEFS}

We define as ``in situ" all stars within Eris' virial radius at the time of analysis that, at the time of formation, were bound only to the main host. Their progenitor gas particles 
became bound to the parent galaxy either by smooth accretion or following the tidal and ram-pressure gas stripping of satellite systems. Stars within Eris' virial radius that formed 
in self-bound subhalos either prior or after infall are called ``ex situ" (labeled below as {\it pre-infall} or {\it post-infall}, respectively).  At the time of analysis, 
ex-situ stars can still be bound to a surviving satellite or be the stellar debris of a disruption event (labeled below as {\it satellite} or {\it smooth}, respectively). This classification largely ensures that the in-situ and ex-situ stellar channels reflect the kinematics and chemical enrichment properties of the main host and its satellite population, respectively. Some ambiguity still remains in the case of stars that form out of gas that was just stripped from infalling satellites \citep[see also][]{Puchwein:2010, Font:2011} and that has not yet mixed with the 
surroundings. 
We assign to the ex-situ (post-infall) category those stars that formed from gas stripped less than 150 Myr earlier; a longer interval between gas stripping and star formation results in in-situ stars instead. This minimal time cutoff is sufficient to prevent those stars -- which would otherwise be tagged ``in-situ'' -- to exhibit phase-space properties similar to the satellite gas they originate from, resulting in intriguing yet not-necessarily realistic {\it in-situ stellar streams} and {\it in-situ stellar shells} reaching galactocentric distances of hundreds of kpc. 
We have tested a series of choices and found that our results are converged for cutoff times larger than about 100 Myr between gas stripping and star formation.\\


\section{Build-Up of Eris' Stellar Component} \label{SEC_ASSEMBLY}

At redshift 0, Eris is a late-type spiral galaxy of virial radius $R_{\rm vir}=235$ kpc and total stellar mass $M_*=3.9\times 10^{10}\,\MSUN$; the latter is comparable to the Milky Way's 
$M_*=4.6 \pm 0.3\times 10^{10}\,\MSUN$ recently determined by \cite{Bovy:2013}. The simulated system is resolved with $N_{\rm DM}=7\times10^6$, $N_{\rm gas}=3\times10^6$, and 
$N_*=8.6\times10^6$ dark matter, gas, and star particles, respectively. Its rotation curve has a value at 8 kpc (the solar circle) of $V_{c,\odot}=206\,\KMS$, in good agreement 
with the recent determination of the local circular velocity, $V_{c,\odot}=218\pm 6\,\KMS$, by \citet{Bovy:2012}. Its total surface mass density, measured within a height of 1.1 kpc 
above and below the disk plane at the Sun location, is 48 $\MSUN$pc$^{-3}$, remarkably consistent with the range of local surface densities recently derived by \cite{Bovy:2013}.
Mock $i-$band images show that Eris has an $i$-band absolute magnitude of $M_i=-21.7$, an extended stellar disk of exponential scale length $R_d=2.5$ kpc, a pseudobulge with S\'ersic
index $n=1.4$, and a bulge-to-disk ratio B/D=0.35 \citep{Guedes:2011}. As shown in \cite{Guedes:2013}, a disk-like star-forming structure is already in place at $z\sim4$, and by $z\sim3.5$ 
it has already grown massive enough to dominate the angular momentum distribution of the stellar component. The present-day $u-r = 1.52$ mag color of Eris is consistent with the 
colors of galaxies that host pseudobulges \citep{Guedes:2013}. Eris's surface brightness profile shows a downbending break at approximately five disk scale lengths, as observed in 
many nearby spiral galaxies \citep{Pohlen:2006}. Eris falls on the Tully-Fisher relation, on the locus of the $\Sigma_{\rm SFR}$-$\Sigma_{\rm HI}$ plane occupied by nearby spiral 
galaxies, and on the stellar mass-halo mass relation at $z=0$.  The predicted correlations of stellar age with spatial and kinematic structures are in good qualitative agreement with 
the correlations observed for mono-abundance stellar populations in the Milky Way \citep{BirdJ:2013}. Twin simulations of the Eris halo that include metal-dependent radiative cooling 
at high temperature ($T > 10^4$ K) and the explicit diffusion of heat and metals between SPH particles have been shown to reproduce quantitatively the properties of the circum-galactic 
medium of galaxies  at $z\sim3$ \citep{Shen:2012a,Shen:2013}. 

\begin{figure}
\includegraphics[width=8cm]{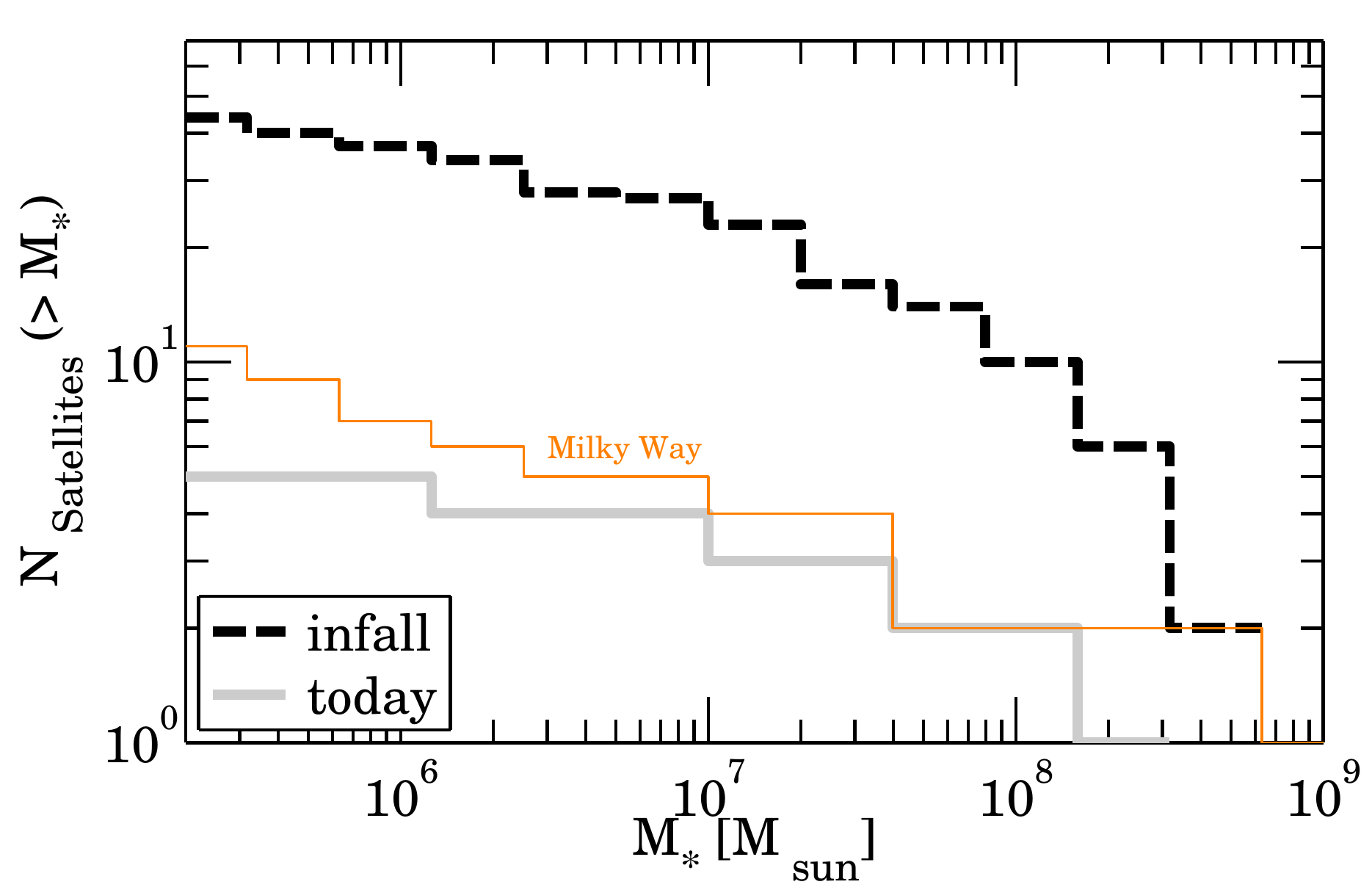}
\caption{Cumulative stellar mass function of Eris satellites within $\RVIR$ at infall ({\it black-dashed curve}) and today ({\it grey solid line}). The orange curve denotes the 
stellar mass function of ``classical" Milky Way dwarf satellites \citep{McConnachie:2012}.}
\label{FIG_SATELLITEFUNCTION}
\end{figure}

Figure \ref{FIG_MWDENSITY_EVOLUTION} depicts the evolution from redshift 4 to today of the stellar density field in Eris, projected in a cube of 240 comoving kpc on a side. 
The three columns compare the contributions from all the stars, the ex-situ stars, and the in-situ stars, using a fixed color scheme that ranges from $10^2$ to $10^8~ \MSUN$ kpc$^{-2}$ in 
surface mass density. At every epoch, the fossil records of past accretion events and mergers are clearly seen in the two left-most columns, at distances between ten and $\gta$ a hundred 
kpc from the center: incompletely phase-mixed material appears as stellar streams, shells, plumes, tidal tails, and self-bound satellite systems \citep[e.g.][]{MartinezDelgado:2010}. 

\begin{figure}
\includegraphics[width=8cm]{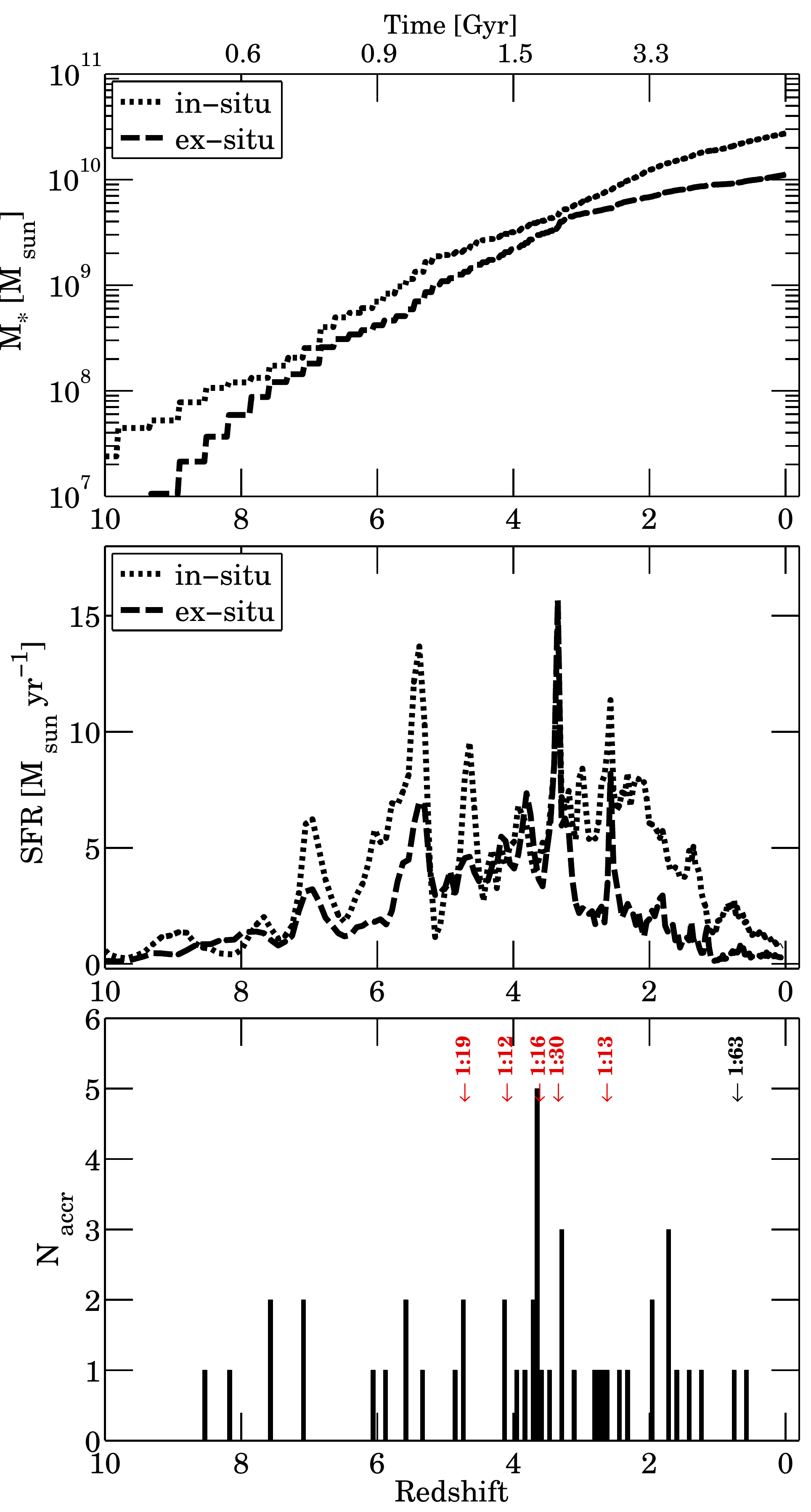}
\caption{The stellar mass assembly history of Eris. Top panel: Stellar mass as a function of redshift for all in-situ and ex-situ stars identified within the 
virial radius at $z=0$.  Middle panel: Star formation rates. Bottom panel: Time distribution of Eris accretion events, including only luminous satellites. 
The red labels marks all the $z<5$ luminous mergers with mass ratios larger than 1:30. An accretion event below $z=1$ is also marked with the 
black label (mass ratio) ``1:63'' as it is among the 5 most luminous satellites ever fallen onto Eris.}
\label{FIG_ASSEMBLYHISTORY}
\end{figure}

The cumulative stellar mass function of Eris satellites is plotted in Figure \ref{FIG_SATELLITEFUNCTION}. Today, there are only five luminous satellites in Eris, 
with maximum circular velocities ranging from 8 to 33 km s$^{-1}$ and stellar masses ranging from $9 \times 10^5$ to $2 \times 10^8\,\MSUN$ in stellar mass. These are the only 
survivors of approximately 50 luminous subhalos at infall. Perhaps because of the particular assembly history and/or the star formation and feedback prescriptions,
Eris shows a deficit of faint satellites relative to the distribution of ``classical" luminous Milky Way's dwarfs \citep[see, e.g.,][]{McConnachie:2012}, a limitation worth keeping in mind 
when interpreting our results.\footnote{\scriptsize{A lower number of surviving satellites in Eris, however, does not necessarily imply an underestimated or overestimated ex-situ fraction 
compared to the Milky Way's, as this also depends on the number, mass and star-formation efficiency of satellites ever fallen in onto the Milky Way, for which no observational constraints are available.}}

Details of the star formation history of Eris are given in Figure \ref{FIG_ASSEMBLYHISTORY}. The top panel shows the assembly history of all in-situ 
and ex-situ star particles identified today within the virial radius. Approximately 70\% of the total stellar mass formed in situ, and two thirds of the ex-situ stars 
formed after infall, i.e. in satellites that are orbiting or merging with the parent halo. Ex-situ stars that form before infall are associated with 10 main 
accretion events occurring between redshift 4 and 0.7. The delay between subhalo accretion and stellar stripping is typically 2 billion years at all times after redshift 
2.5. In the second and third panels of Figure \ref{FIG_ASSEMBLYHISTORY}, we report the rates of star formation of in-situ and ex-situ stars and the number of accreted luminous 
satellites, respectively. In-situ star formation dominates at all redshifts, but for a lapse of time between $4.5 \lta z \lta 3$ that is characterized by repeated and relatively 
massive accretion events (see bottom panel). The simulated galaxy is currently forming stars at approximately 1 $\MSUN$ yr$^{-1}$. The in-situ star formation rate is 
sustained above a level of 5 $\MSUN$ yr$^{-1}$ at all times $5 \lta z \lta 2$, with peaks of star formation above 15 $\MSUN$ yr$^{-1}$ at $z = 5.4,~ 3.3, $ and 2.6. 
Interestingly, two of these peaks occur in correspondence to two major accretion events. Star formation in Eris' progenitor satellites is less bursty and sizeable 
compared to the main halo: star formation rates in the dwarf population range from 2 to 5 $\MSUN$ yr$^{-1}$ between $z\sim 6$ and $z \sim 2$. 

\begin{table}
\begin{center}
\caption{Eris stellar masses and fractions to total at $z=0$, decomposed in the different star-formation modes.}
\label{TAB_STARMASSES}
\begin{tabular}{l c c}
\hline
&&\\
					& Stellar Mass 			& Fraction to Total\\
					& [$\MSUN$] 			& [\%]\\
&&\\					
\hline
&&\\
All					& $3.9\times 10^{10}$	& 100		\\
In-situ 		 		& $2.7\times 10^{10}$ 	& 71.1		\\
Ex-situ 				& $1.1\times 10^{10}$ 	& 28.9		\\
&&\\
Ex-situ (pre-infall)          	& $2.9\times 10^{9}$ 	& 7.5			\\
Ex-situ (post-infall) 	& $8.2\times 10^{9}$ 	& 21.4		\\
&&\\
Ex-situ (smooth) 		& $1.1\times 10^{9}$ 	& 27.4		\\
Ex-situ (satellites) 		& $5.6\times 10^{8}$ 	& 1.5		\\
&&\\
\hline
\end{tabular}
\end{center}
\end{table}

Table \ref{TAB_STARMASSES} summarizes the mass budget among Eris different stellar components at the present epoch. About a quarter of the ex-situ post-infall stars have formed out of gas that was stripped too recently from infalling satellites to have undergone chemical and phase mixing with the surroundings (see Section \ref{SEC_DEFS}).
The stellar accretion history of Eris reflects the criteria with which its host halo was selected from a sample of candidates: a Milky Way-size halo
with a particularly quiescent merger history. Although dark subhalos are being accreted also at low redshifts, the last luminous satellite fell onto Eris
at $z = 0.54$, 5 billion years ago, with a total mass ratio of 1:180. The most recent merger with mass ratio $>$ 1:20 occurred at $z = 2.66$; this is also the most massive halo
ever accreted by the parent host, with $M_{\rm vir}=1.8\times 10^{10}\,\MSUN$ and $M_*=6.3\times 10^{8}\,\MSUN$ at infall. This massive system is responsible for more than
35 per cent of the ex-situ stars formed after infall. Before getting completely destroyed by $z \sim 1$, it deposits more than $10^9~ \MSUN$ in stars
within the innermost regions of Eris. Eris does not undergo any major $>$ 1:4 merger during its whole assembly history. At $z<5$, we count only 5
mergers with mass ratios $>$ 1:30. These are all luminous satellites with infall at $z>2.5$.

\begin{figure*}
\begin{center}
\includegraphics[width=16cm]{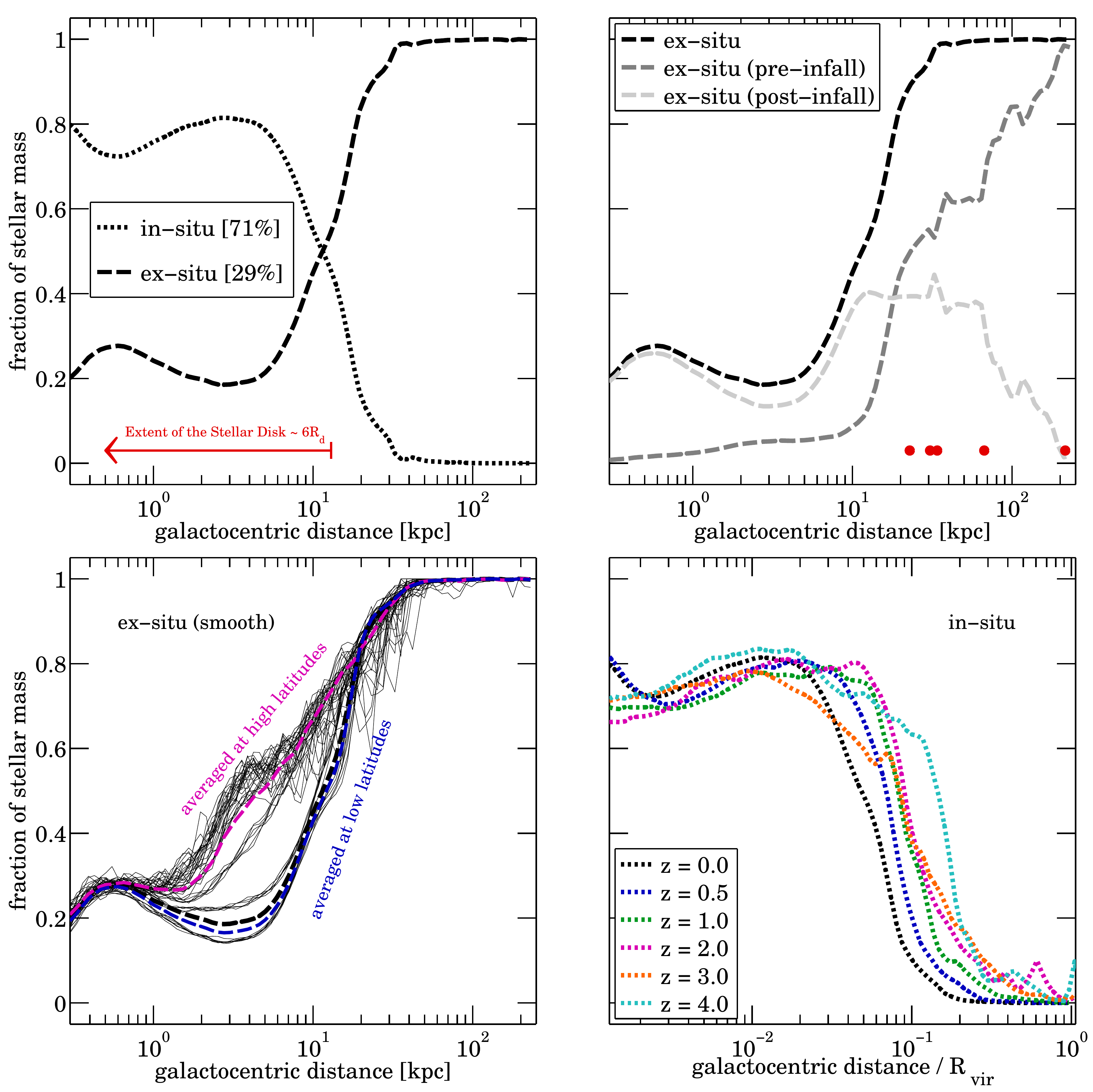}
\caption{Mass fraction in spherical shells of in-situ and ex-situ stars as a function of galactocentric distance. Top left panel: Spherically-averaged 
stellar mass fractions of in-situ (dotted line) and ex-situ (dashed line) stars at $z=0$. The label marks the total mass fraction over the whole halo, 71\% for in-situ stars and 29\% 
for ex-situ stars. Top right panel: Same for all (black line), pre-infall (dark grey line) and post-infall (light grey line) ex-situ stars. 
The red filled circles denote the locations of Eris' surviving satellites at $z=0$.
Bottom left panel: Same for all ex-situ stars not in satellites (``smooth") along different lines of sight. The whole sky average (black line) is compared with mass profiles at high 
($>30$ degrees from the disk plane, magenta lines) and low ($<30$ degrees from the disk plane, blue lines) latitudes. The in-situ contribution drops much faster at high latitudes than 
along the stellar disk. Bottom right panel: Same for all in-situ stars at different redshifts and as a function of galactocentric distance in units of the virial radius $R_{\rm vir}$. 
Note that in the two bottom panel we have removed the contribution of satellite stars to the overall normalization so that features along the profiles at fixed redshift reflect the 
spatial properties of the smooth stellar component.
}
\label{FIG_FRACTPROFILES}
\end{center}
\end{figure*}

\section{Fractions of In-Situ and Ex-Situ Stars} \label{SEC_ACROSSMW}

In Figure \ref{FIG_FRACTPROFILES}, we show the relative contributions of in-situ and ex-situ stars to the total stellar mass as a function of galactocentric distance and 
cosmic time. The top left panel gives the fractional mass in spherical shells of in-situ and ex-situ stars at $z=0$, including stars in satellites. Although in-situ stars 
dominate the overall stellar mass budget, ex-situ stars prevail at large radii. The transition zone between in-situ and ex-situ dominance is narrow and occurs at about 10 kpc 
from the center. At radii larger than 30 kpc, the contribution of in-situ stars drops below 5\% in mass. Normalized mass profiles for pre- and post-infall ex-situ stars 
are depicted in the top right panel, together with the location of surviving luminous satellites today. The bump in the ex-situ component within the innermost 1 kpc region 
comes from stars brought into the bulge by a massive luminous satellite that continued forming stars until redshift 1 while spiralling down after infall, and that was ultimately disrupted. 
There is a second bump in the contribution of ex-situ post-infall stars at distances $\gta 30$ kpc that is associated with the two most massive satellites surviving today and 
visible in the density maps of Figure \ref{FIG_MWDENSITY_EVOLUTION} (the other satellites are too small to leave a prominent signature in the mass profiles).  
The two subhalos are characterized by 10-billion-years prolonged star formation histories, with about $2\times 10^9 ~\MSUN$ in stars formed and subsequently stripped after infall.
The contribution of post-infall ex-situ stars undergoes a sharp drop at distances larger than 60-70 kpc for two main reasons: 1) the bulk of such stars are formed within a handful 
of the most massive satellites that quickly spiral in toward the center owing to dynamical friction and that thereby leave less sizable debris at large distances; and 2) bursts of star formation 
in such massive satellites appear to occur following their pericenter passages, never larger than a few tens of kpc.

To quantify the lumpiness and asymmetry of the stellar halo visible by eye in Figure \ref{FIG_MWDENSITY_EVOLUTION}, we show in the bottom left panel of Figure \ref{FIG_FRACTPROFILES}
the profiles of ex-situ stars along lines of sight at different azimuthal and polar angles relative to Eris' stellar disk. The observer has been located at the galaxy center. 
Owing to the presence of the stellar disk, the in-situ to ex-situ balance is strikingly different at distances smaller than $\sim 20$ kpc, with the in-situ (ex-situ) contribution 
decreasing (increasing) much faster at high than at low latitudes: the transition between in-situ and ex-situ stars occurs at about 3-5 kpc from the center along directions perpendicular 
to the disk. Spherical symmetry does not hold even at galactocentric distances larger than $\sim 15$ kpc: the scatter and variety in the line-of-sight profiles at high latitudes 
demonstrate that fine-grained structures such as stellar streams and shells are also visible in the ex-situ vs. in-situ balance. Finally, the bottom right panel depicts 
the spherically-averaged mass profiles of in-situ stars at different redshifts. Note how the in-situ fractional contribution can be several times larger at early times than 
today for distances greater than 0.02 $\RVIR$. The transition radii where ex-situ stars start dominating the mass profile are different at different redshift, yet always 
in the range between $\sim$ 10 and 20 comoving kpc at $z<4$. We expand on these features in the next section.

\begin{figure}
\includegraphics[width=8cm]{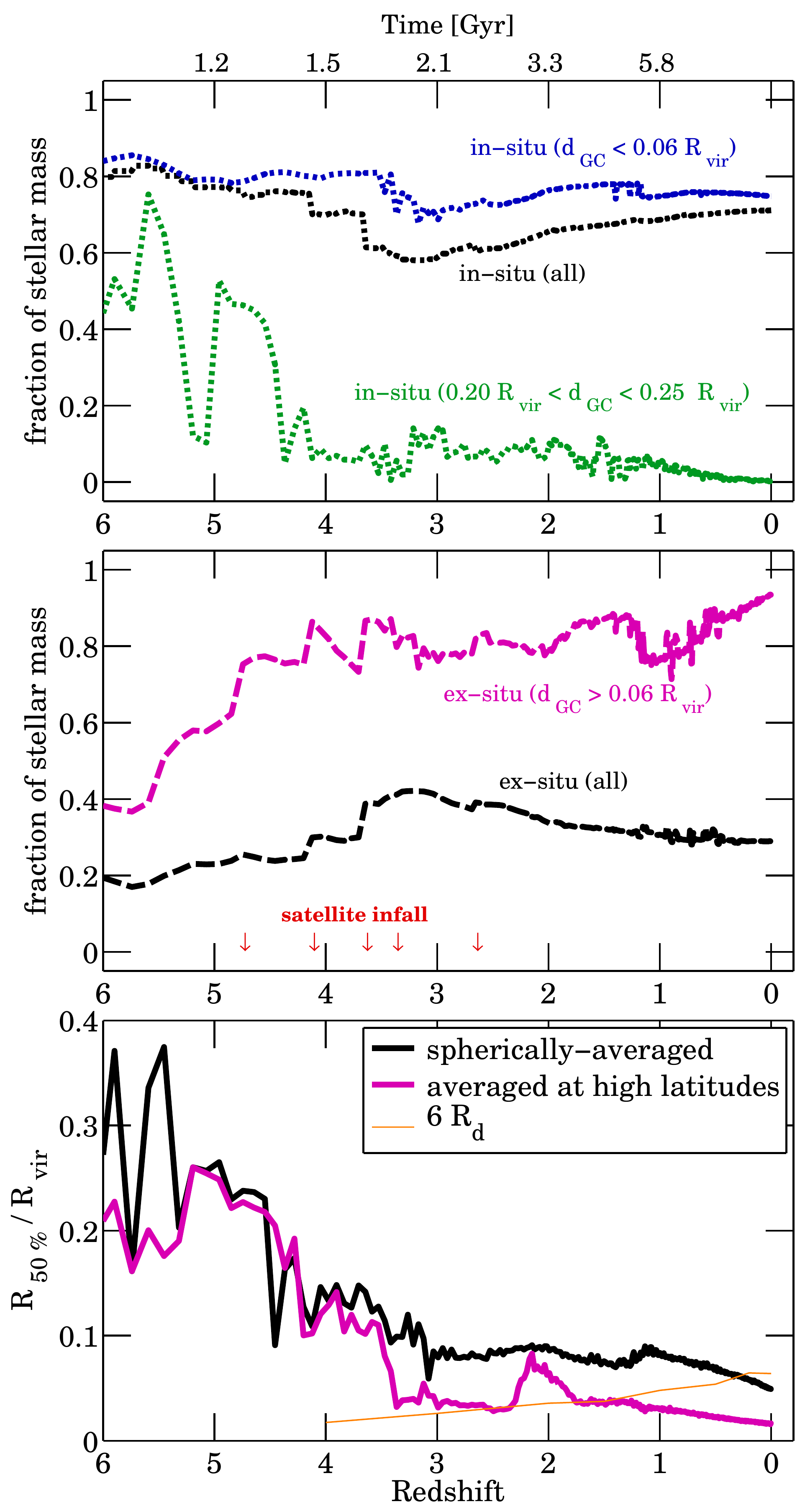}
\caption{Top panel: Evolution of the in-situ mass fraction as a function of redshift within the halo (dotted black curve), within the inner 0.06$\RVIR$ (i.e. 15 kpc or 6 disk scale lengths 
at $z=0$, blue dotted curve), and across a spherical shell between 0.2 and 0.25$\RVIR$ from the galaxy center (green dotted line). Middle panel: Evolution as a function of redshift 
of the ex-situ mass fraction within the halo (black dashed curves) and within the volume beyond 0.06$\RVIR$. 
The red arrows mark the more important mergers characterizing Eris' assembly history. Bottom panel: Radius (in units of $\RVIR$) where the in-situ and ex-situ stellar densities are equal, 
as a function of redshift. For reference, Eris' disk scale length \citep{Guedes:2013}, multiplied by 6, is shown by the solid thin orange line.}
\label{FIG_FRACTEVOLUTIONS}
\end{figure}

\subsection{Time Evolution} \label{SEC_EVOLUTION}

Figure \ref{FIG_FRACTEVOLUTIONS} tracks the evolution of the global in-situ mass fraction (top panel), ex-situ mass fraction (middle panel), and of the transition radius 
between in-situ and ex-situ domination (bottom panel), as a function of redshift. In what follows, we adopt an evolving boundary equal to $0.06\,\RVIR$ to 
separate stars belonging to the ``inner'' and ``outer'' regions of the galaxy: at $z=0$, such boundary corresponds to 15 kpc, the extent of Eris' stellar 
disk (about 6 times the disk scale 
length). The in-situ mass fraction remains between 60 and 80\% of the total at all epochs, with a shallow minimum around the epoch of merger activity at $z \sim 3$, when the 
ex-situ fraction is maximized. Note how, beyond the extent of the $z=0$ stellar disk, the contribution of ex-situ stars reaches a local minimum around $z\sim 1.1$: this is 
the time when the most massive satellite finally merges with the main host, after having repeatedly perturbed its innermost regions for the preceding $\sim 10^9$ years. 
After $z\sim 1$, the ex-situ outer component steadily grows up to the current $\sim$90 percent for all stars at $d_{\rm GC}>0.06\,\RVIR$, 
during the long, recent period of relatively quiet accretion history when no satellite --luminous nor dark-- plunges at small impact parameters towards Eris' center. 
Conversely, the contribution of in-situ stars to the outer regions of Eris was as large as 20-25 percent, several times higher 
than their current balance today. The picture presented so far, where the ex-situ to in-situ balance in different regions of the galaxy is modulated by the merger 
history of the host, suggests the possibility that in-situ stars may be displaced by several kpc from their birth sites during the early assembly history of 
Eris.\footnote{Note that the displacement of disk stars due the presence of spiral density waves (radial 
migration, either inwards and outwards) is the manifestation of a different phenomenon than the one we are witnessing here: although certainly in place, radial migration 
can justify displacements only by up to the extent of the stellar disk. Moreover, spiral-induced resonances associated with radial migration cannot alone justify the 
thickening and heating of a stellar disk \citep[see][]{VeraCiro:2014}, let alone displacements by many kpc beyond the disk scale-height of Eris.}

\begin{figure}
\begin{center}
\includegraphics[width=8.5cm]{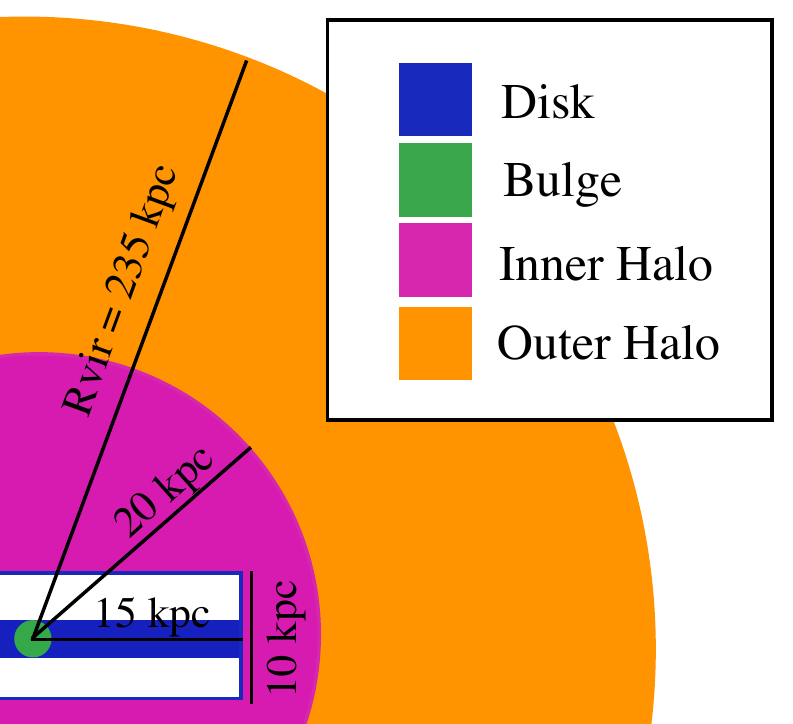}
\caption{Schematic illustration of the simple spatial decomposition adopted in this analysis for Eris. See text for details.}
\label{FIG_DECOMPOSITION}
\end{center}
\end{figure}

\begin{table*}
\begin{center}
\caption{Stellar masses, in-situ and ex-situ fractions in each major stellar component.}
\label{TAB_FRACTIONS}
\begin{tabular}{lccccc}
\hline
&&&&&\\
                                                & Milky Way                                             & Disk                                          & Bulge                                         & Inner Halo                            & Outer Halo    \\
&&&&&\\
\hline
&&&&&\\
Total stellar mass                      & $3.9\times 10^{10}\,\MSUN$                    & $1.9\times 10^{10}\,\MSUN$            & $1.5\times 10^{10}\,\MSUN$            & $7.8\times 10^{8}\,\MSUN$     & $1.7\times 10^{9}\,\MSUN$\\
&&&&&\\
In-situ                         & 71 \%                                                         & 77 \%                                                 & 76 \%                                                 & 25 \%                                         & 3\%\\
Ex-situ                         & 29 \%                                                 & 23 \%                                                 & 24 \%                                                 & 75 \%                                         & 97\%\\
Ex-situ (satellites)    & 1.5 \%                                                        & -                                                     & -                                                     & -                                             & 21\%\\

&&&&&\\
\hline
\end{tabular}
\end{center}
\end{table*}

Finally, in the bottom panel of Figure \ref{FIG_FRACTEVOLUTIONS}, the evolution of the transition scale between in-situ and ex-situ stars (the radius where the local 
densiy of in-situ stars drops below 50\% of the total) is given as a function of redshift, in units of Eris' time-varying virial radius. This transition scale 
is shown in the case of stellar densities that 
are spherically-averaged across the whole sky, and at high galactic latitudes from Eris's stellar disk. This is well defined since $z\sim 3.5$ and is aligned to today's 
stellar plane since $z\sim 2$, when it temporarily flips by almost 90 degrees (causing the bump in the magenta curve). Below redshift 2, the transition radius between in-situ 
and ex-situ stars remains approximately constant. After the last satellite sinks into the central regions of the system at $z\sim 1$, the transition radius decreases slowly 
from 7\% of the virial radius to today's 5\%, as in-situ stars displaced from their birth sites at early times appear to fall back towards the center of Eris. 

\subsection{In-Situ and Ex-Situ By Components} \label{SEC_DECOMPOSITION}

We complete our description of the interplay between ex-situ and in-situ stars by providing their relative contributions within the different morphological stellar components: 
disk, bulge, inner, and outer halo. We opt for a simple spatial decomposition rather than one based on stellar kinematics, chemical properties, or photometry. 
Such decomposition is easily reproducible by observers and allows us to predict potential trends in the bulk properties of the Milky Way's components without being biased 
a priori towards specific regions of the stellar-property space.

The adopted decomposition is schematically illustrated in Figure \ref{FIG_DECOMPOSITION}. The coordinate system is defined by the angular momentum of Eris' stellar disk, identified 
by the orbital circularity as in \cite{Guedes:2011, Guedes:2013}: this is the only kinematically derived quantity. A sphere of 1.5 kpc radius around the center of mass of Eris' halo 
is termed the {\it bulge}. The stellar {\it disk} is a cylinder of $\pm 1.5$kpc thickness and 15 kpc radius: it includes the so-called thin and thick disks and excludes the bulge.
To avoid interpretation difficulties, we exclude from our analysis a cylindrical, 10 kpc-thick region around the disk. All the volume beyond such ``transition" region is called the {\it halo}.
Following \cite{Carollo:2010}, we distinguish between the {\it inner halo} within 20 kpc from the center and the {\it outer halo} beyond: we do so to study possible trends at large
galactocentric distances. As seen above, the separation between an inner and an outer halo cannot be motivated in Eris by the transition from an in situ-dominated to an ex situ-dominated 
regime, as this occurs at 11 kpc (spherically-averaged) and 5 kpc (along directions perpendicular to the disk). Our decomposition is a good proxy of more sophisticated 
kinematic$+$photometric decompositions, which yields in Eris a disk scale height and disk length at $z=0$ of 490 pc and 2.5 kpc, respectively \citep{Guedes:2011, BirdJ:2013}. 
The mass of the so-defined outer stellar halo, excluding satellites, is $\sim 1.3 \times 10^9 \MSUN$, in good agreement with observational estimates of the Galactic 
stellar halo \citep[of order $10^9 \MSUN$][]{Morrison:1993,Bell:2008,Deason:2012a,Rashkov:2013}.

As detailed in Table \ref{TAB_FRACTIONS}, 87 percent of the stellar mass in Eris resides within the bulge and the disk, whereas the stellar halo accounts for only 6 percent of the 
total. More than one fifth of the disk and bulge stellar content is composed by ex-situ stars (23 and 24 percent respectively). Conversely, ex-situ stars make approximately 75 and 97 
percent of the inner and outer halo, respectively. Interestingly, there are at present four times more ex-situ mass in the disk and in the bulge than in the halo. 
A quarter of the inner halo is composed of in-situ stars that have been displaced from their original birth, even though these account for only a few times $10^8\MSUN$ in stellar mass. Finally, about 20 percent of the outer stellar halo mass is enclosed in five self-bound satellites with infall redshift $\lesssim$ 1.6.\\

\section{Stellar Properties and \\Observational Signatures} \label{SEC_PROPERTIES}

As they depend on the star-formation history of the main host, in-situ stars are expected to span a variety of ages, metallicities, and kinematic properties. Similarly, the properties of 
ex-situ stars mirror the star-formation histories of satellite dwarfs before and after infall. In this section we measure such properties in Eris' stars and discuss them in light of 
recent observations of the Milky Way stellar disk and halo. We characterize each star in the simulation by the following quantities: galactocentric distance, galactocentric radial velocity, formation redshift or age, iron abundance, 
and orbital circularity. The orbital energy of a star is closely connected to its radial velocity, while the orbital circularity encodes information about orbital angular momentum and eccentricity. 
Here, we define a star's circularity $\epsilon$ as the ratio between its angular momentum component along the stellar disk axis and the angular momentum it would have if it was on a circular 
orbit at the given radius, i.e., $\epsilon\equiv J_z/J_c(r_\star)$.
For a rotationally-supported stellar component (as the thin disk), the circularity distribution peaks at $\epsilon = 1$; a positive circularity denotes an orbit corotating 
with the disk, while a negative value implies counter rotation.

\begin{figure*}
\begin{center}
\includegraphics[width=16cm]{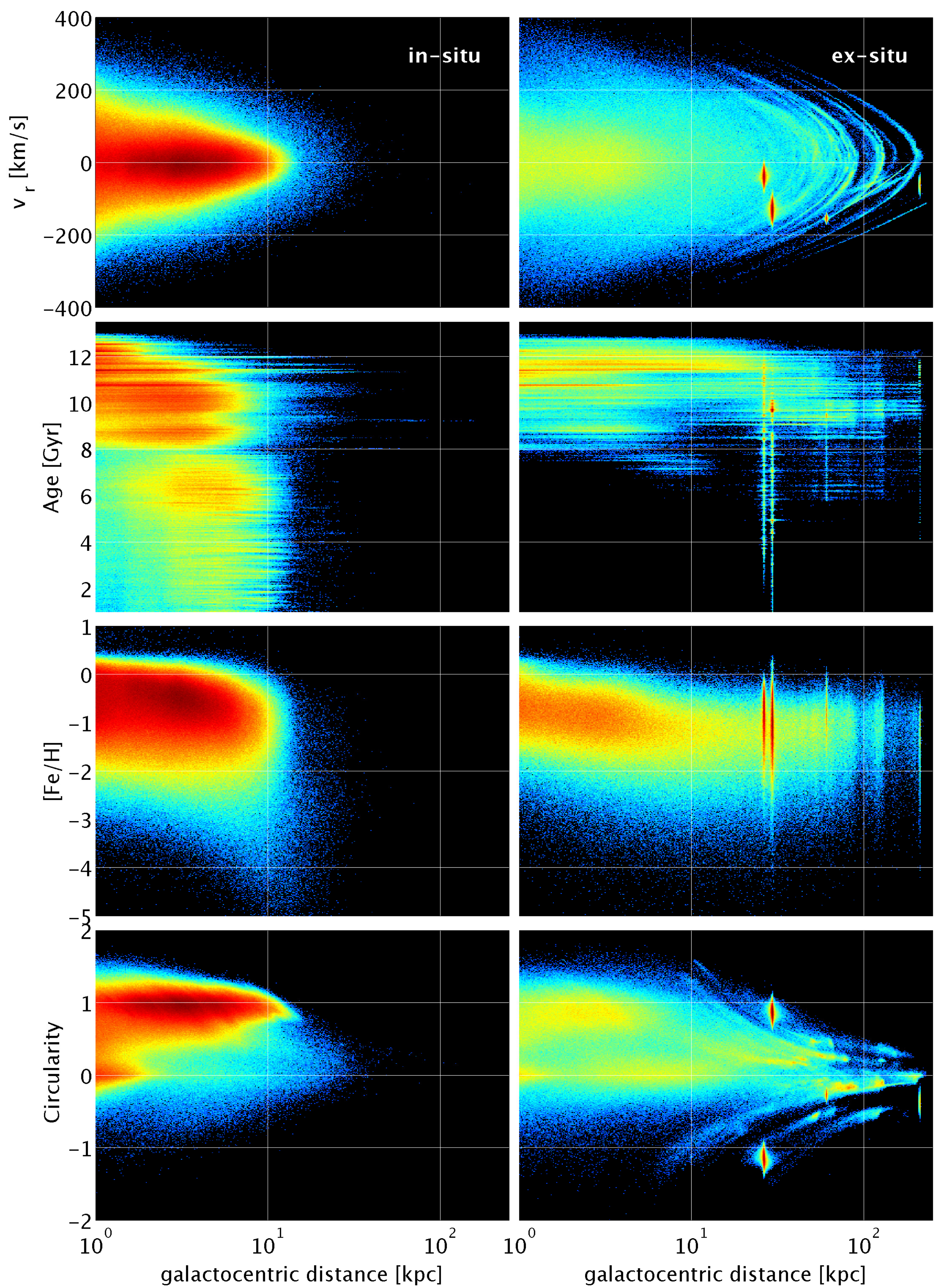}
\caption{Present-day distributions of in-situ (left panels) and ex-situ (right panels) stars in Eris. From top to bottom: distribution in the 
radial velocity-distance, age-distance, metallicity-distance, and circularity-distance plane, color-coded according to stellar densities
per logarithmic bin in distance. The ex-situ category includes stars in self-bound satellites, readily visible as localized overdensities at a fixed distance.
}
\label{FIG_PROPERTIES}
\end{center}
\end{figure*}
 
In Figure \ref{FIG_PROPERTIES}, the aforementioned stellar characteristics are shown for in-situ and ex-situ stars, as a function of galactocentric distance, 
color-coded according to stellar densities per logarithmic bin in distance. The radial velocity-distance plot for ex-situ stars (top right panel) is a 
typical phase-space diagram \citep[e.g.][]{Bullock:2005}, which reveals surviving satellites as well as long-lived fine-grained structures 
such as shells and streams (in configuration space). A population of ``hot" ex-situ stars can be identified in the disk, at small galactocentric distances: these are 
stars brought in corotating, circular orbits by the last significant merger at redshift 1 (``ex-situ disk"), which is also responsible for the formation of a light dark-matter disk in Eris \citep{Pillepich:2014aa}.

Ex-situ stars are typically older than in-situ, with a 
median age of 11.3 Gyr compared to 9.8 Gyr. The age distribution of ex-situ stars is broadened towards young ages by star formation that occurs in satellites after infall. 
Distant ``smooth" halo stars have median ages that {\it decline} from 11.2 Gyr at 20 kpc (inner halo) to 9.8 Gyr at 70 kpc (outer halo).  The median age in Eris' inner halo is in close agreement with the age of the inner halo of the Milky Way, 11.4 $\pm$ 0.7 Gyr \citep{Kalirai:2012}, whereas the same Milky Way's observations seem to suggest an even older outer stellar halo. 
However, the mean ages we measure in the whole inner and outer halos (10.3 and 9.9 Gyr, respectively) are actually indistinguishable in light 
of their large standard deviations, 2.6 and 1.8 Gyrs respectively. At fixed (large) galactocentric distance, in-situ stars have median ages that are 
systematically younger than their ex-situ neighbors, by about 2.5-3 Gyr. This observation per se does not support the popular picture of a younger, inner halo 
dominated by in-situ stars, but may provide some insight on the mechanisms at play.

All the surviving satellites in our simulation are characterized by extended star formation histories, with durations
spanning from about 3 to about 10 Gyrs: these can be seen as vertical stripes in the second panel from the top on the right (Figure \ref{FIG_PROPERTIES}). None of the satellites is currently star forming, 
but the two most massive surviving dwarfs at $z=0$ have stellar populations as young as 1 or 2 Gyrs. Although it is possible that the adopted numerical SPH scheme 
may artificially underestimate the impact of ram-pressure stripping on the gaseous content of orbiting subhalos \citep[see e.g.][]{Sijacki:2012}, we note here that our simulated 
satellites appear broadly consistent with resolved stellar population studies of Local Volume dwarfs \citep{Weisz:2011, Weisz:2014}, which also show the signatures of 
extended star-formation activity. 

The in-situ metallicity distribution peaks at [Fe/H] $= -0.5$, while its median vary from [Fe/H] $= -0.4$ in the innermost regions of the galaxy, to about [Fe/H] $=-1.3$ at 
10 kpc. Whereas the metallicity trend and slope is in qualitative agreement with observations of disk galaxies, we note that the simulated abundances are systematically lower by 
0.5 in [Fe/H] than the measured metallicities in the Galaxy's disk \citep[e.g.][ from SEGUE data]{Bovy:2012b, Cheng:2012b}. This is likely caused by the steep 
\citet{Kroupa:1993} IMF adopted for the Eris run, which produces IMF-averaged metal yields that are a factor of 3 smaller compared to more recent determinations. Ex-situ stars 
are more metal-poor than in-situ, with a distribution that peaks at [Fe/H] $= -0.9$ and is less skewed towards high abundances. The inner and outer halos of Eris have mean iron 
abundances of [Fe/H] $=-1.5$ and [Fe/H] $\sim -1.3$, respectively, with broad distributions. This appears partially in contrast with halo stars measurements by, e.g., 
\cite{Carollo:2007} or \cite{An:2013}, who identify a negative metallicity trend from the inner, [Fe/H] $\sim -1.6$, to the outer, [Fe/H]$\sim -2.0$ halo. 
It is unclear at this stage whether the discrepancy is caused by our (non-standard) definition of inner halo or by a different assembly history between Eris and the Milky Way.
We note, however, that our metallicity profiles in the stellar halo are consistent with the results of other numerical works, where metallicity gradients in the galaxy outskirts 
vary with the accretion history \citep{Tissera:2013, Tissera:2014}, but where median iron abundances are never lower than $-1.5$ when averaged over the whole volume of the outer halo 
\citep[e.g.][]{Font:2006,Font:2011}. 

Finally, the bottom panels of Figures \ref{FIG_PROPERTIES} depict the distributions of circularity. Eris in-situ stars features a sizeable rotationally-supported disk-like structure 
($\epsilon = 1$) with a subdominant non-rotating bulge ($\epsilon = 0$). At distances $\gta 10$ kpc, in-situ stars are distributed rather symmetrically around $\epsilon=0$, with 
only a modest hint of prograde rotation: from their kinematics properties only, they appear rather indistinguishable from the background of ex-situ stars at similar locations.	
Structures in the circularity-distance plane of ex-situ stars are more complex, with both satellites and debris stars retaining the orbital and phase-space properties of the 
associated accretion event. Satellites are accreted both in a corotating as well as in a counter rotating fashion relative to the stellar disk. 
At small distances, the ex-situ distribution is strongly shifted towards positive circularities because dynamical friction is more efficient at dragging towards the 
disk plane incoming satellites that are on prograde orbits relative to the disk.\\

\begin{figure}
\begin{center}
\includegraphics[width=8.5cm]{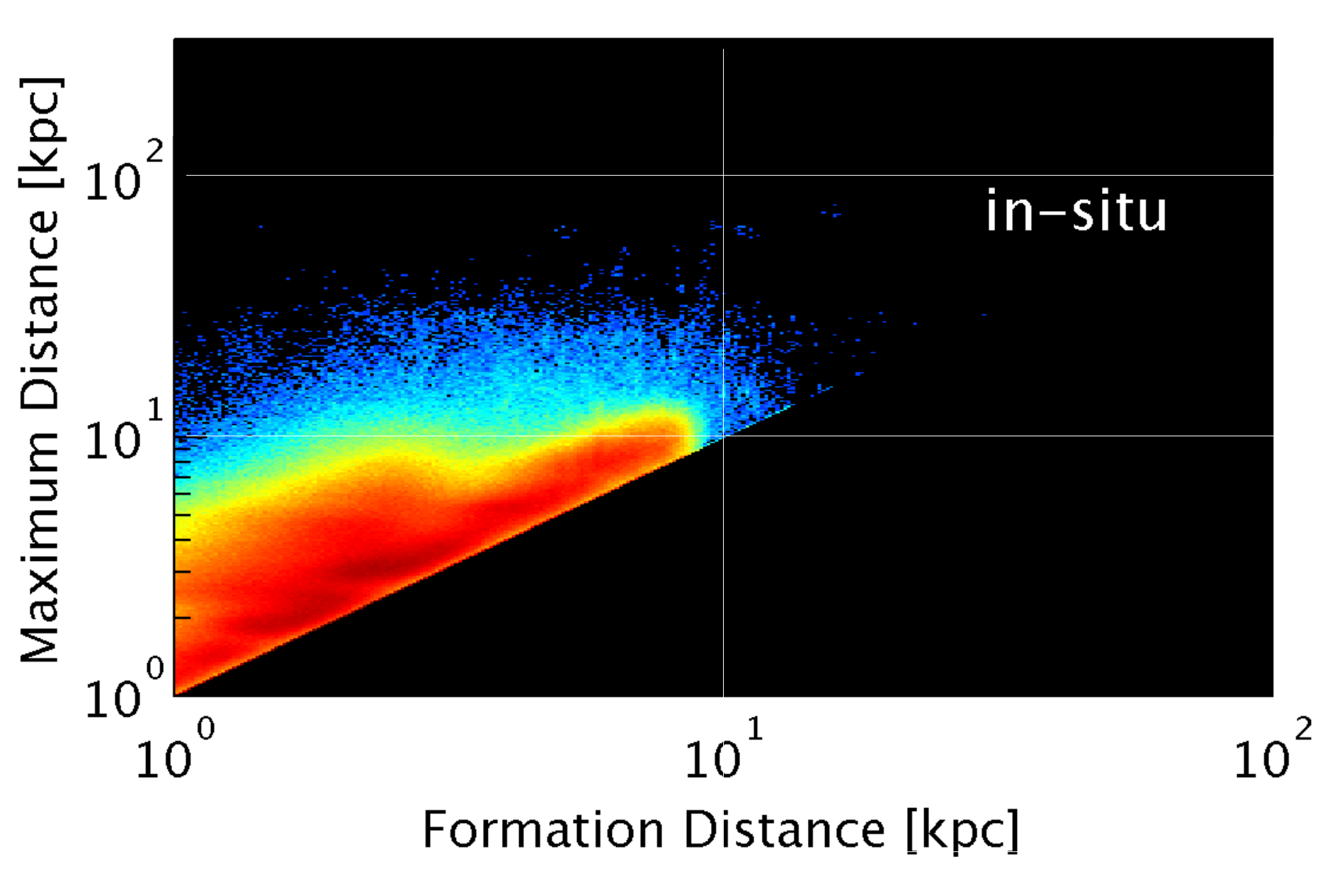}  
\caption{Maximum distance from the center of Eris reached by in-situ stars across their lifetimes vs. their distance at formation. Units are in physical kpc, color coded according to the stellar densities per logarithmic bin in distance (dark red corresponds to more than $10^3$ stars per squared bin with $\Delta({\rm log}_{10}d[kpc]) = 0.006$).
}
\label{FIG_DISTANCES}
\end{center}
\end{figure}

\section{Discussion}
\label{SEC_DISC}

In the previous sections, we have shown two interesting features of the stellar spatial distribution in Eris: 1) in-situ stars, which formed within the innermost regions of the galaxy, 
can be found far in the stellar halo, tens of kpc away from their original birth sites; 2) ex-situ stars, formed within satellite galaxies before or after infall, can be found today in the disk 
and bulge of the main host. The same mergers which bring ex-situ stars towards the baryonic disk are also responsible for the displacement of in-situ stars into the halo. While the details of the heating process 
will be addressed in another paper, Figure \ref{FIG_DISTANCES} demonstrates that many in-situ stars effectively travel large distances from their formation sites during their lifetimes. 
There we show that in-situ stars indeed formed in the highest density regions of the simulated galaxy, i.e. within 10-12 physical kpc away from the galactic center (about 5-6 times the scale length of Eris' stellar disk today). Moreover, our analysis suggests that some in-situ stars have traveled during their lifetime to even larger galactocentric distances than where we find them today: more than $10^9\MSUN$ in-situ stars have traveled to distances $\gtrsim 12$ physical kpc during Eris assembly history, even though this number drops to about $10^8\MSUN$ for maximum distances larger than 20 physical kpc.\\ 


\section{Summary} \label{SEC_FINAL}

We have analyzed the assembly history of the stellar components in ``Eris", a cosmological N-body+SPH simulation of a close Milky Way analog. The combination of high mass, spatial, and time resolution 
allows a detailed study of the relative contributions of in-situ star formation within the main host and ex-situ star accretion from satellite galaxies. 
The classification adopted for this study is meant to ensure that the chemical and physical properties of the two categories naturally reflect the characteristics of the ISM out of which they have formed.
We have identified and kept track of different sub-categories of ex-situ stars, including a smooth component consisting of tidally-striped stars, those still belonging to surviving satellites at the 
present epoch, stars formed before vs. those formed after infall.

We have investigated the balance between the in-situ and ex-situ channels, as a function of galactocentric distance and among the different morphological stellar components of the galaxy.
Our main results with Eris can be summarized as follows:

\begin{itemize}

\item Approximately 70 percent of today's stars formed in-situ, and more than two thirds of the ex-situ stars formed within satellites {\it after} infall (Table \ref{TAB_STARMASSES});

\item Both in-situ and ex-situ stars have undergone sizable galactocentric displacements during their lifetimes: the majority of Eris' ex-situ stars are found today in the disk and in the bulge, whereas approximately 25\% of the inner ($\lesssim 20$ kpc) halo is composed of in-situ stars that have been displaced from their original birth sites during Eris' early assembly history (Table \ref{TAB_FRACTIONS});

\item The stellar halo -- which by our definition accounts for only 6 percent of the total stellar mass -- is dominated by ex-situ stars, whereas in-situ stars dominate the mass profile at distances $\lta 5$ kpc from the center at high latitudes (Figure \ref{FIG_FRACTPROFILES});

\item The transition radius where the local density of in-situ stars drops below 50 percent of the total appears to be only mildly dependent on the merger history of the host halo, and it measures $\lesssim$ 5 percent of the virial radius at all times after redshift $3-4$ (Figure \ref{FIG_FRACTEVOLUTIONS});

\item Across the whole galaxy, ex-situ stars are on average 1.5 Gyr older and more metal poor than in-situ stars: their distributions are broad, and skewed towards the young and metal-rich tails by those stars formed within orbiting satellites with {\it extended} star formation histories (Section \ref{SEC_PROPERTIES});

\item The circularity distribution of ex-situ and in-situ stars plotted as a function of current distance from the center encodes a wealth of information: the stellar halo in Eris appears a multicomponent ensemble with both smooth and localized phase-space structures in both prograde and retrograde orbits with respect to the stellar disk (Figure \ref{FIG_PROPERTIES}).

\end{itemize}

Our analysis of Eris' assembly and star-formation history shows that infalling satellites spanning the whole mass-ratio range contribute to the build-up of the stellar halo, whereas the more massive satellites are responsible for both 
depositing ex-situ stars within the innermost regions of the galaxy and for displacing in-situ stars from their original birth sites. Additional work is required to understand the physical mechanism responsible for the lifting of 
in-situ disk stars into the halo, and to quantify the impact on our findings of 1) different halo masses, 2) diverse, more violent merger histories, and 3) alternative implementations of the star formation efficiencies in dwarf galaxies at high redshifts, for which observations are still ambiguous. 

\acknowledgments
Support for this work was provided by the NSF through grants OIA--1124453 and AST--1229745, and by NASA through grant NNX12AF87G.
The Eris Simulation was carried out at NASA's Pleiades supercomputer. AP thanks Alexander Knebe for providing support and input on the
Amiga Halo Finder, and Sijing Shen, Alis Deason, Ryan Cooke, and Connie Rockosi for many stimulating discussions on the topic of this paper.

\bibliography{ErisProjects_Bibliography}
\bibliographystyle{apj}

\end{document}